\documentclass[journal]{IEEEtran}
\usepackage{caption}
\usepackage{times}
\usepackage{epsfig}
\usepackage{graphicx}
\usepackage{amsmath}
\usepackage{amssymb}
\usepackage{comment}
\usepackage{multirow}
\usepackage{subcaption}
\usepackage{cuted}
\usepackage{capt-of}
\usepackage[table]{xcolor}
\usepackage{longtable}
\usepackage{url}
\usepackage{hyperref}
\usepackage{cite}

\usepackage{array,float,xfrac}
\begin{document}
%
\title{OpenEarthMap-SAR: \\A Benchmark Synthetic Aperture Radar Dataset for Global High-Resolution Land Cover Mapping}
%
%
%
\author{Junshi~Xia,~\IEEEmembership{Senior Member,~IEEE,}
Hongruixuan~Chen,~\IEEEmembership{Graduate Student Member,~IEEE,}
Clifford Broni-Bediako,~\IEEEmembership{Member,~IEEE,} Yimin Wei,
Jian~Song, Naoto~Yokoya,~\IEEEmembership{Member,~IEEE}
        
\thanks{This work was supported in part by JSPS, KAKENHI under Grant Number 23K24865, and Next Generation AI Research Center of The University of Tokyo.}
\thanks{Junshi Xia and Clifford Broni-Bediako are with RIKEN Center for Advanced Intelligence Project (AIP), RIKEN, Tokyo, 103-0027, Japan (e-mail: junshi.xia@riken.jp, clifford.broni-bediako@riken.jp).}
\thanks{Hongruixuan Chen, Yimin, Wei and Jian Song, and Naoto Yokoya are with Graduate School of Frontier Sciences, The University of Tokyo, Chiba, Japan and  ith RIKEN Center for Advanced Intelligence Project (AIP), RIKEN, Tokyo, 103-0027, Japan (e-mail: qschrx@gmail.com; yimin.wei@riken.jp;song@ms.k.u-tokyo.ac.jp; yokoya@k.u-tokyo.ac.jp).}
}

\maketitle

\begin{abstract}
   High-resolution land cover mapping plays a crucial role in addressing a wide range of global challenges, including urban planning, environmental monitoring, disaster response, and sustainable development. However, creating accurate, large-scale land cover datasets remains a significant challenge due to the inherent complexities of geospatial data, such as diverse terrain, varying sensor modalities, and atmospheric conditions. Synthetic Aperture Radar (SAR) imagery, with its ability to penetrate clouds and capture data in all-weather, day-and-night conditions, offers unique advantages for land cover mapping. Despite these strengths, the lack of benchmark datasets tailored for SAR imagery has limited the development of robust models specifically designed for this data modality.
To bridge this gap and facilitate advancements in SAR-based geospatial analysis, we introduce OpenEarthMap-SAR, a benchmark SAR dataset, for global high-resolution land cover mapping. OpenEarthMap-SAR consists of 1.5 million segments of 5033 aerial and satellite images with the size of 1024$\times$1024 pixels, covering 35 regions from Japan, France, and the USA, with partially manually annotated and fully pseudo 8-class land cover labels at a ground sampling distance of 0.15--0.5 m. We evaluated the performance of state-of-the-art methods for semantic segmentation and present challenging problem settings suitable for further technical development. The dataset also serves the official dataset for IEEE GRSS Data Fusion Contest Track I. The dataset has been made publicly available at \textcolor{magenta}{\url{https://zenodo.org/records/14622048}}.

\end{abstract}


%
\IEEEpeerreviewmaketitle

\section{Introduction}
\label{sec:1}
\par High-resolution Synthetic Aperture Radar (SAR) imagery plays a crucial role in land cover mapping, especially for applications like land use planning \cite{Shermeyer2020SpaceNet}, disaster response \cite{Adriano2021Learning, chen2025bright}, and resource management \cite{Michele2023High}. While medium-resolution satellite imagery, e.g., 10-30m ground sampling distance (GSD),  has been widely used in global land cover mapping \cite{Chini2018Towards, Hafner2022Unsupervised}, high-resolution SAR (sub-meter GSD) provides significant advantages, especially in detecting features like buildings, roads, and other infrastructure in challenging conditions, such as adverse weather or night-time.

\par Recent developments in SAR-based land cover mapping have seen major progress, with several studies successfully mapping building footprints over large areas using high-resolution SAR imagery \cite{Shermeyer2020SpaceNet, Michael2024Deep}. Unlike optical imagery, SAR has a unique set of characteristics that make it particularly valuable for geospatial mapping and monitoring, especially in regions where optical imagery faces significant limitations. One of the most notable advantages of SAR is its insensitivity to cloud cover and weather conditions. SAR's all-weather capability means it can provide continuous and consistent monitoring over long periods, which is crucial for tracking dynamic changes in the environment, such as urban growth, deforestation, or disaster response. This consistency is particularly valuable for real-time monitoring applications, as it reduces the data gaps typically associated with optical imagery when clouds or adverse weather block the view~\cite{TSOKAS2022117342,10612244,9351574}.

\par With the rise of deep learning technologies, substantial efforts have been made to create benchmark datasets for high-resolution SAR image analysis \cite{Shermeyer2020SpaceNet, Adriano2021Learning, Li2022MCANet, chen2025bright}. Datasets providing large-scale SAR data have significantly advanced the development of object detection methods, supporting key applications like vehicle detection and infrastructure monitoring. Benchmark datasets such as SIVED~\cite{rs15112825}, SADD~\cite{9761751}, MSAR~\cite{wang2023sar}, and SARDet-100K~\cite{li2024sardet100k} have been crucial in driving this progress. Additionally, SAR datasets have been used for change detection and disaster damage mapping \cite{Adriano2021Learning, chen2025bright}, further demonstrating their real-world applicability.



\begin{figure*}[!t]
    \centering
    \includegraphics[height=3.50in]{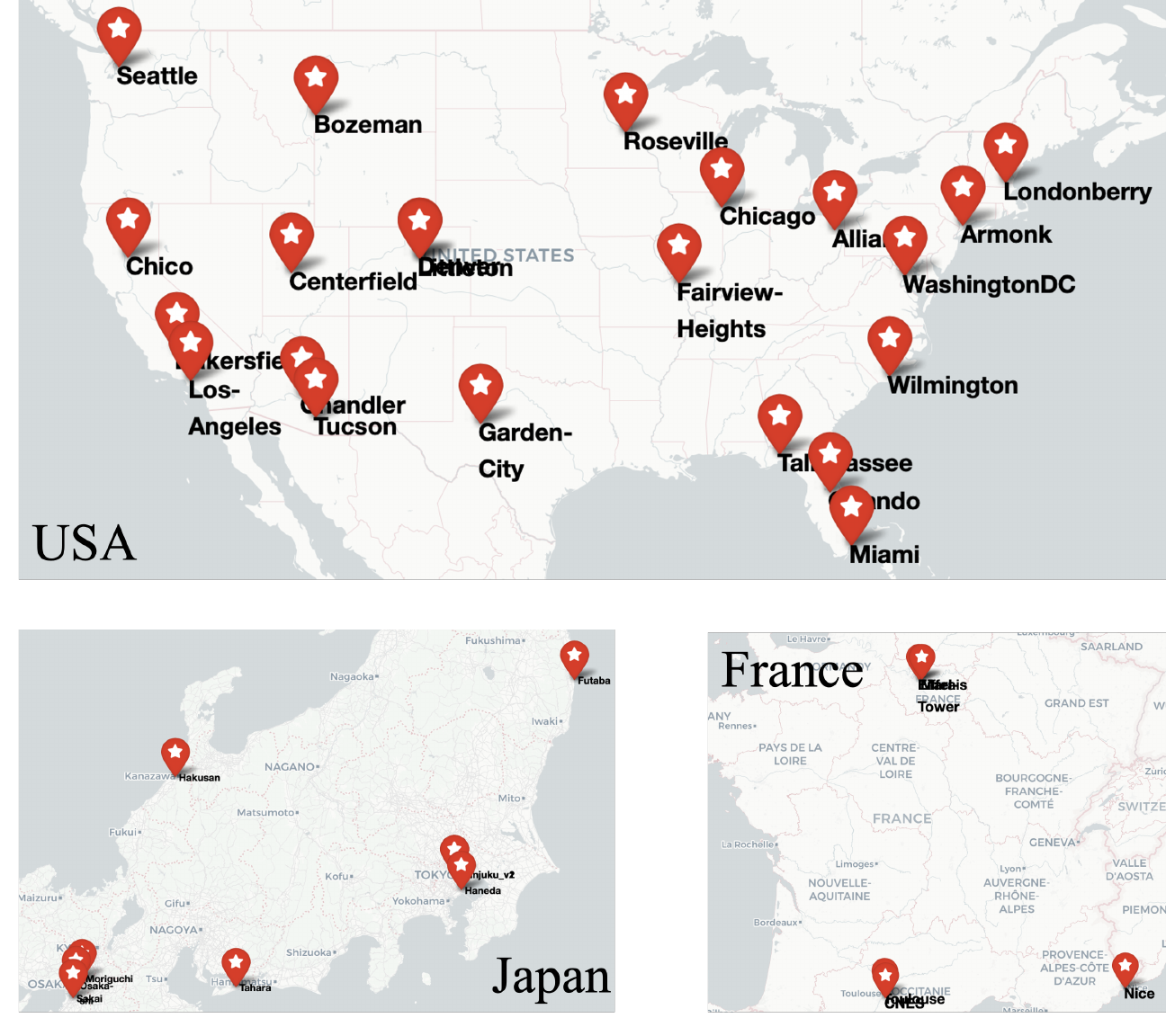}
   \caption{Geographic locations of 35 regions present in this work. }
  \label{fig:location}
\end{figure*}

\begin{table*}[t!]
\caption{Comprehensive Overview of High-Resolution SAR Benchmark Datasets for Segmentation-related task: Building Extraction (BE), Land Cover Mapping (LC), and Change Detection (CD).
}
\label{tab:comparison}
\vspace{-1mm}
\begin{center}
\scalebox{1.2}{
\begin{tabular}{c c c c c c c }
\hline\hline
Image level & GSD (m) & Dataset & Task & Classes & Countries & Regions \\

\hline
\multirow{3}{*}{Meter level} & 3 & FU-SAR~\cite{9369836} & LC & 4 & 1 & 8  \\
& 1 & GF-3 Building~\cite{9444638} & BE & 2 & 7  & 9  \\
& 1.2-3.3 & BDD~\cite{ADRIANO2021132} & CD & 3 & 8 & 9  \\
\hline
\multirow{3}{*}{Sub-meter level} & 0.5 & SpaceNet 6~\cite{9150641} & BE & 2 & 1 & 1   \\
& 0.35 & QQB~\cite{sun2023qqb} & CD & 2 & 1 & 1  \\
& 0.3-1 & BRIGHT~\cite{chen2025bright} & CD & 3 & 12 & 21  \\

\cline{3-7}
& 0.15--0.5 & OpenEarthMap-SAR & LC & 8 & 3 & 35  \\
\hline\hline
\end{tabular}
}
\vspace{-7mm}
\end{center}
\end{table*}

\par The challenge with benchmark datasets for semantic segmentation at sub-meter SAR resolution lies in the limited availability of high-quality public datasets, poor annotation quality, and the difficulties associated with detailed annotation. Publicly available datasets are scarce due to the lack of high-resolution SAR imagery in many regions and the commercial restrictions on redistributing such data. Moreover, the cost and complexity of manually annotating sub-meter resolution SAR images are substantial, requiring significant expertise and resources, making it difficult to achieve consistent, reliable, and high-quality annotations. As a result, many datasets resort to using coarse annotations, such as bounding boxes or minimal class labels, or rely on indirect sources like OpenStreetMap, which may not fully capture the complexity of the land cover. The inherent challenges of SAR sensing mechanisms further complicate direct annotation. SAR images have limitations in terms of the visibility of certain land cover classes due to their sensitivity to surface roughness, moisture, and other factors, meaning that some classes, such as roads and cropland, might not be detectable or accurately delineated. Even identifying specific features such as individual buildings  can be challenging, as SAR images often lack the level of detail found in optical imagery. 
Thus, ensuring high-quality, accurate annotations for SAR images remains a significant technical hurdle, requiring innovative solutions to overcome the challenges of annotation quality and data availability.

\par To address these challenges, we introduce OpenEarthMap-SAR, a new addition to the OpenEarthMap~\cite{xia2023openearthmap} series, designed as a benchmark dataset for global high-resolution land cover mapping using SAR imagery. OpenEarthMap-SAR represents a significant advancement over existing datasets in both geographic diversity and annotation quality (Table~\ref{tab:comparison}). The dataset includes 8-class pseudo land cover labels and partial real labels at a ground sampling distance (GSD) of 0.15-0.5 meters, encompassing 5033 images from 35 regions across three continents: Europe, America, and Asia, with specific coverage in France, Japan, and the USA. The 8 classes in the OpenEarthMap-SAR dataset are bareland, rangeland, developed space, road, tree, water, agricultural land, and building. The pseudo labels for these classes are generated using pre-trained OpenEarthMap models \cite{xia2023openearthmap, yokoya2024submeter, broni2024generalized, xia2023generating, Chen2024ObjFormer}.

\par We evaluate the performance of baseline models for semantic segmentation across different modalities: Optical, SAR, and their combination (SAR+Optical), with results presented for varying settings: pseudo labels, real labels, and the combination of both pseudo and real labels. We identify key challenges in applying these methods to SAR data, highlighting areas for technical improvement. This initiative focuses on developing robust methods for land cover mapping in all weather conditions using SAR data, emphasizing its potential for improving accuracy and reliability in real-world mapping scenarios.


\par

\section{The Dataset}\label{sec:2}
\subsection{Data preparation}\label{sec:2.1}
\par In this work, we have carefully selected 35 diverse regions from the USA, Japan, and France (Fig.~\ref{fig:location}), encompassing a wide range of urban and rural landscapes. These regions were chosen to represent various geographic and environmental conditions, ensuring that the dataset captures the complexity and diversity of real-world land cover scenarios.

\par The SAR datasets are sourced from Umbra Space~\footnote{\url{http://umbra-open-data-catalog.s3-website.us-west-2.amazonaws.com}}, featuring Spotlight mode with a resolution ranging from 0.15m to 0.5m. The optical datasets consist of aerial images are from the National Agriculture Imagery Program (NAIP)~\footnote{\url{https://datagateway.nrcs.usda.gov}}, the French National Institute of Geographic and Forest Information (IGN)~\footnote{\url{https://geoservices.ign.fr/bdortho}}, and the Geospatial Information Authority of Japan (GSI)~\footnote{\url{https://maps.gsi.go.jp/development/ichiran.html}}. The optical data comprises red, blue, and green bands, whereas the SAR data primarily contains amplitude information in the VV or HH polarization bands.The optical datasets were converted from digital numbers to reflectance values and subsequently standardized to an 8-bit format. For the SAR imagery, pre-processing was conducted by the data provider, followed by conversion to an 8-bit format. Despite both the optical and SAR images being geocoded, some misalignment remained between them. To rectify this, multiple experts manually aligned the paired optical and SAR datasets, cross-checking their work to ensure they met the required standard.

\begin{figure*}
    \centering
    \includegraphics[width=1\linewidth]{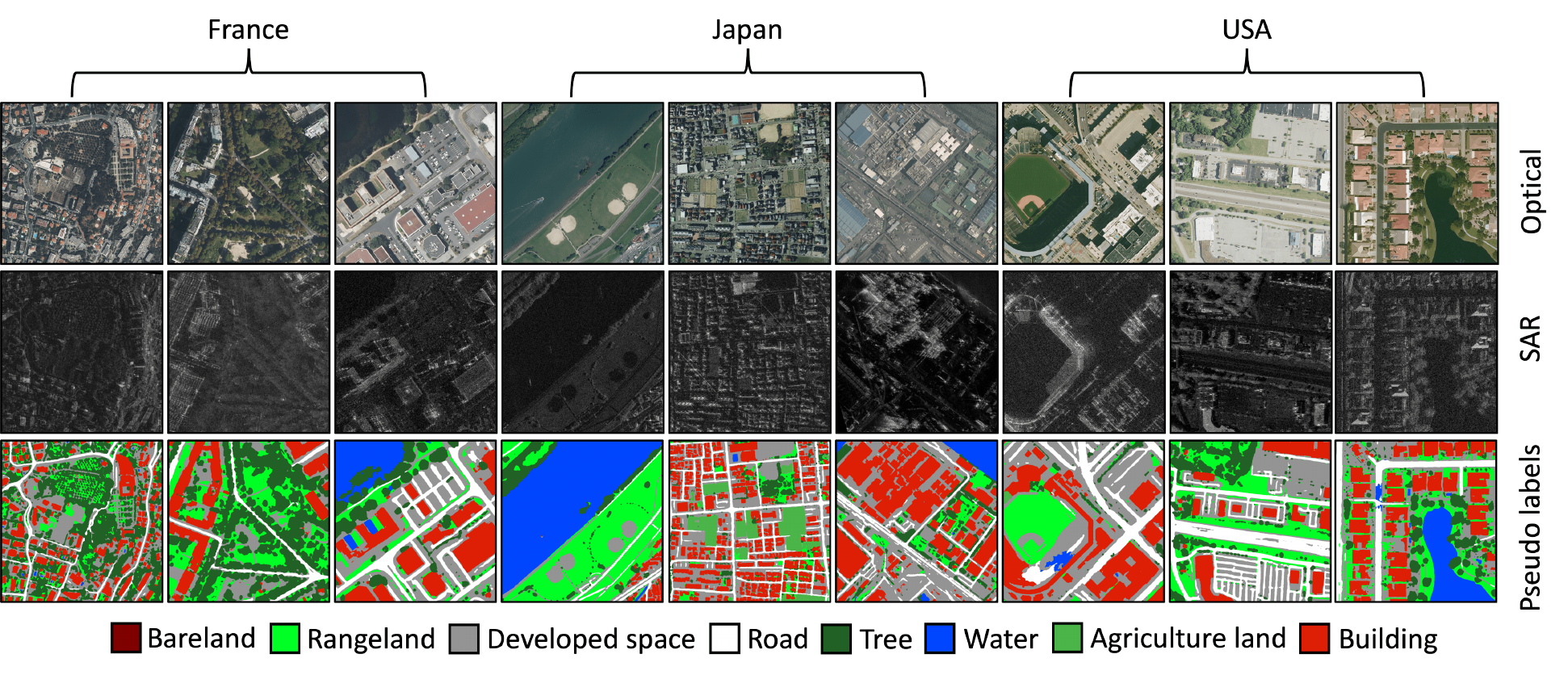} \\
    \caption{Examples of datasets. {SAR image © 2024 Umbra Lab, Inc., used under \href{https://creativecommons.org/licenses/by/4.0/}{CC BY 4.0 license}. Optical images of the 1-3 columns © 2024 National Institute of Geographic and Forest Information (IGN), France, used under \href{https://creativecommons.org/licenses/by/2.0/}{CC BY 2.0 license}; Optical image of the 4-6 columns courtesy of Geospatial Information Authority of Japan (GSI); Optical images of the 7-9 columns courtesy of the National Agriculture Imagery Program (NAIP), USA.} 
    }
    \label{fig:examples}
\end{figure*}

\subsection{Classes, Annotations, and Data Split}\label{sec:2.2}

\definecolor{bareland}{rgb}{0.50196078,0,0}
\definecolor{rangeland}{rgb}{0,1,0.14117647}
\definecolor{develop}{rgb}{0.58039216,0.58039216,0.58039216}
\definecolor{road}{rgb}{1,1,1}
\definecolor{tree}{rgb}{0.13333333,0.38039216,0.14901961}
\definecolor{water}{rgb}{0,0.27058824,1}
\definecolor{agriculture}{rgb}{0.29411765,0.70980392,0.28627451}
\definecolor{building}{rgb}{0.87058824,0.12156863,0.02745098}

\begin{table}
\caption{The number and proportion of pixels and the number of segments of the eight classes. }
\label{tab:oem_pixels_segments}
\vspace{-2mm}
\begin{center}
\scalebox{1}{
\begin{tabular}{c c c c c}
\hline\hline
Color & \multirow{2}{*}{Class} & \multicolumn{2}{c}{Pixels}& Segments\\
\cline{3-4}
(HEX) & & Count (M) & (\%) & (K) \\
\hline
\colorbox{bareland}{\textcolor{white}{800000}} & \textit{Bareland}  & 85   & 1.6 & 8.7 \\
\colorbox{rangeland}{\textcolor{white}{00FF24}}& \textit{Rangeland} & 1263 & 23.9 & 392.3 \\
\colorbox{develop}{\textcolor{white}{949494}}& \textit{Developed space}  & 1222  & 23.1 & 451.2 \\
FFFFFF & \textit{Road}   & 415  & 7.9 & 69.4 \\
\colorbox{tree}{\textcolor{white}{226126}} & \textit{Tree} & 755  & 14.3 & 342.1\\
\colorbox{water}{\textcolor{white}{0045FF}}& \textit{Water} & 543  & 10.3 & 19.3 \\
\colorbox{agriculture}{\textcolor{white}{4BB549}}& \textit{Agriculture land} & 344  & 6.5 & 9.3 \\
\colorbox{building}{\textcolor{white}{DE1F07}}& \textit{Building} & 640  & 12.1 & 178.5 \\
\hline\hline
\end{tabular}
}
\vspace{-4mm}
\end{center}
\end{table}

\noindent\textbf{Classes:} We provide annotations and pesudo labels with eight classes: \textit{bareland}, \textit{rangeland}, \textit{developed space}, \textit{road}, \textit{tree}, \textit{water}, \textit{agriculture land}, and \textit{building}. The class selection is consistent with our previous benchmark datasets, OpenEarthMap \cite{xia2023openearthmap} as well as other benchmark datasets (e.g., LoveDA~\cite{wang2021loveda} and DeepGlobe~\cite{demir2018deepglobe}) with sub-meter GSD. Table~\ref{tab:oem_pixels_segments} presents the number and proportion of labeled pixels, as well as the number of segments for each class, derived from the pseudo labels. OpenEarthMap-SAR contains more pixels and segments for Rangeland and Developed space, while tree, agricultural land and building are more abundant in OpenEarthMap~\cite{xia2023openearthmap}.



\noindent\textbf{Annotations: } The pseudo labels are generated from pre-trained OpenEarthMap models \cite{xia2023openearthmap, yokoya2024submeter, broni2024generalized, xia2023generating, Chen2024ObjFormer}. In addition, for each region, we manually annotated 20 images to ensure the quality and consistency of the dataset, thereby improving the accuracy of the labels.


\begin{table}[ht]
\centering
\caption{Agreement and IoU between pseudo and real labels. }
\scalebox{1.5}{
\begin{tabular}{lcc}
\hline
\textbf{Class}           & \textbf{Agreement}   & \textbf{IoU}   \\
\hline
\textit{Bareland}        & 0.1063        & 0.0211         \\
\textit{Rangeland}       & 0.7827        & 0.4705         \\
\textit{Developed Space} & 0.6959        & 0.5251         \\
\textit{Road}            & 0.7038        & 0.6129         \\
\textit{Tree}            & 0.6899        & 0.6413         \\
\textit{Water}           & 0.9396        & 0.8483         \\
\textit{Agriculture Land}& 0.6692        & 0.5809         \\
\textit{Building}        & 0.8708        & 0.7812         \\
\hline
\textbf{Mean}            & \textbf{0.6823} & \textbf{0.5602} \\
\hline
\end{tabular}
}
\label{tab:classwise_metrics}
\end{table}

\par Additionally, when assessing the performance using pseudo-labels and 700 real labels, the mean agreement rate between the two sets of annotations drops to 68$\%$ (seen in Table~\ref{tab:classwise_metrics}), especially for the class \textit{Bareland} and \textit{Agriculture Land}. 
This lower agreement indicates a noticeable gap in the quality of pseudo-labels compared to manually-labeled ones. The discrepancy arises because pseudo-labels are typically generated by models, which may not capture all the complex variations present in high-resolution remote sensing images. These models may also struggle with generalization across different image regions, leading to misclassifications and inconsistencies in the labeling process. In contrast, human annotators can use contextual understanding and local knowledge to improve accuracy, although their performance can still vary due to the inherent complexity of remote sensing imagery. The 68\% agreement highlights the ongoing challenges in achieving high-quality annotations from automated systems, especially compared to the near-perfect results achieved by human annotators under ideal conditions. In the end, we collected a total of 5033 images with the size of 1024$\times$1024 pixels from 35 regions from USA, Japan and France (seen in the examples in Fig.~\ref{fig:examples}).

\noindent\textbf{Data split: }
The dataset is split into several subsets for different tasks, with distinct training, validation, and testing configurations. For the \textbf{semantic segmentation task}, the models will be evaluated using 490 images from the manually labeled dataset, which contains 14 real labels per region. The following training scenarios are defined:

\begin{itemize}
    \item \textbf{Using only pseudo labels}: The model will be trained exclusively on 4333 images, respectively, all containing pseudo labels.
    \item \textbf{35 real labels (1 per region) + pseudo labels}: The model will be trained using 35 real labels (1 per region) combined with pseudo labels from the unannotated images.
    \item \textbf{175 real labels (5 per region) + pseudo labels}: The model will be trained using 175 real labels (5 per region) combined with pseudo labels from the unannotated images.
    \item \textbf{Real labels (1/5 per region) only}: The model will be trained exclusively on the real labels (1 or 5 per region).
\end{itemize}




For the \textbf{semantic segmentation task}, we will evaluate the performance with \textbf{SAR}, \textbf{Optical}, and \textbf{Optical+SAR} data combinations to assess the models across different modalities.

\subsection{Comparison with other datasets}
\par The OpenEarthMap-SAR dataset stands out in comparison to other high-resolution SAR benchmark datasets in terms of its region coverage and class diversity. While datasets like FU-SAR~\cite{9369836}, GF-3 Building~\cite{9444638}, and BDD~\cite{ADRIANO2021132} focus on specific tasks such as Land Cover Mapping (LC), Building Extraction (BE), and Change Detection (CD) with smaller class counts, OpenEarthMap-SAR offers a broader range of classes (8 in total) and covers more regions (35), spanning across 3 countries. It also presents a resolution range of 0.15–0.5 meters, placing it in the sub-meter level, similar to datasets such as SpaceNet 6~\cite{9150641} and BRIGHT~\cite{chen2025bright}. However, the OpenEarthMap-SAR dataset surpasses other datasets in terms of its extensive coverage, both geographically and in the number of regions, making it a valuable resource for studying land cover patterns in SAR imagery. Despite the smaller number of classes compared to certain other datasets like FU-SAR, which includes 4 classes, OpenEarthMap-SAR's diverse geographic scope and multiple region coverage provide a unique advantage for comprehensive land cover analysis.

\begin{table*}
\caption{IoU and mIoU performance of baseline models with Optical, SAR, and SAR+Optical modalities across various labeling scenarios: pseudo labels (P), pseudo + 1 real label per region (P+R1), pseudo + 5 real labels per region (P+R5), 1 real label per region (R1), and 5 real labels region (R5).}
\label{tab:rgb_sar_combined_results}
\begin{center}
\scalebox{1}{
\begin{tabular}{c c c c c c c c c c cc}
\hline \hline
Method & Modality     & Case   & Bareland & Grass & Developed & Road  & Tree  & Water & Agriculture Land & Buildings & mIoU (\%) \\
\hline \hline
\multirow{15}{*}{U-Net}  & \multirow{5}{*}{Optical}          & P      & 0.11    & 48.04 & 53.48    & 61.96 & 62.46 & 83.67 & 64.51    & 78.23     & 56.56 \\
       &              & P+R1   & 0.29    & 48.60 & 53.63    & 63.45 & 64.36 & 84.39 & 66.06    & 78.51     & 57.41 \\
       &              & P+R5   & 1.47    & 49.76 & 55.02    & 62.83 & 65.50 & 83.84 & 68.06    & 79.02     & 58.19 \\
       &              & R1     & 0.00    & 43.29 & 41.06    & 54.57 & 69.67 & 67.65 & 41.81    & 66.84     & 48.11 \\
       &              & R5     & 32.63   & 58.35 & 53.72    & 64.02 & 75.95 & 85.21 & 73.07    & 77.89     & 65.10 \\
\cline{2-12}
 & \multirow{5}{*}{SAR}          & P      & 0.00    & 26.89 & 24.50    & 21.51 & 37.41 & 69.04 & 60.50    & 41.14     & 35.13 \\
       &              & P+R1   & 0.09    & 26.51 & 24.75    & 21.39 & 37.31 & 67.07 & 57.95    & 41.85     & 34.61 \\
       &              & P+R5   & 0.16    & 26.94 & 24.52    & 22.86 & 39.70 & 71.62 & 66.70    & 42.25     & 36.84 \\
       &              & R1     & 0.00    & 13.06 & 18.78    & 13.27 & 35.88 & 35.85 & 56.30    & 31.22     & 25.54 \\
       &              & R5     & 13.12   & 24.79 & 23.73    & 19.74 & 43.98 & 51.85 & 54.02    & 39.62     & 33.86 \\
\cline{2-12}
  & \multirow{5}{*}{SAR+Optical}      & P      & 0.09    & 48.30 & 52.51    & 60.22 & 62.92 & 82.55 & 66.84    & 77.13     & 56.32 \\
       &              & P+R1   & 0.30    & 48.33 & 52.99    & 62.69 & 62.07 & 83.23 & 66.66    & 78.02     & 56.79 \\
       &              & P+R5   & 0.00    & 41.07 & 42.14    & 50.90 & 56.68 & 74.80 & 64.84    & 67.92     & 49.79 \\
       &              & R1     & 0.00    & 38.87 & 38.58    & 50.01 & 64.87 & 63.78 & 52.75    & 65.56     & 46.80 \\
       &              & R5     & 13.68   & 50.26 & 52.26    & 61.21 & 74.53 & 86.75 & 66.70    & 75.87     & 60.16 \\
\hline \hline
\multirow{15}{*}{SegFormer}  & \multirow{5}{*}{Optical}       & P      & 0.06    & 49.83  & 53.98    & 61.63 & 64.64 & 82.86  & 67.52    & 77.55     & 57.26 \\
           &           & P+R1   & 0.09    & 49.69  & 54.18    & 64.40 & 63.00 & 84.24  & 68.84    & 77.81     & 57.88 \\
           &           & P+R5   & 0.03    & 50.06  & 54.52    & 64.23 & 65.32 & 83.23  & 65.47    & 78.51     & 58.07 \\
           &           & R1     & 8.83    & 49.30  & 46.60    & 58.30 & 73.91 & 73.52  & 50.39    & 71.75     & 54.07 \\
           &           & R5     & 25.61   & 56.04  & 55.77    & 65.64 & 77.06 & 88.64  & 67.77    & 78.20     & 64.34 \\
\cline{2-12}
           & \multirow{5}{*}{SAR}       & P      & 0.00    & 27.23  & 27.35    & 20.79 & 37.25 & 69.01  & 62.43    & 42.06     & 35.77 \\
           &           & P+R1   & 0.00    & 28.09  & 27.48    & 23.26 & 39.05 & 69.89  & 62.95    & 42.59     & 36.66 \\
           &           & P+R5   & 0.00    & 27.89  & 28.53    & 24.16 & 39.74 & 72.12  & 62.24    & 43.14     & 37.23 \\
           &           & R1     & 0.97    & 21.17  & 16.85    & 15.25 & 40.18 & 43.00  & 53.06    & 34.46     & 28.12 \\
           &           & R5     & 17.16   & 21.47  & 25.10    & 21.17 & 43.67 & 53.77  & 57.47    & 37.20     & 34.63 \\
\cline{2-12}
           & \multirow{5}{*}{SAR+Optical}   & P      & 0.10    & 49.79  & 53.31    & 62.28 & 62.52 & 83.41  & 68.95    & 77.25     & 57.20 \\
           &           & P+R1   & 0.16    & 53.62  & 54.51    & 63.92 & 70.50 & 84.23  & 71.64    & 77.92     & 59.56 \\
           &           & P+R5   & 0.49    & 50.21  & 55.09    & 64.26 & 64.88 & 87.03  & 68.14    & 78.46     & 58.57 \\
           &           & R1     & 6.55    & 47.38  & 44.09    & 54.01 & 71.66 & 70.73  & 50.44    & 69.49     & 51.79 \\
           &           & R5     & 38.19   & 56.46  & 54.93    & 64.16 & 76.79 & 87.68  & 70.54    & 76.74     & 65.69 \\
\hline \hline

\multirow{15}{*}{VMamba}      & \multirow{5}{*}{Optical}       & P      & 0.00  & 47.95  & 52.54    & 62.76 & 64.54 & 83.34  & 63.24    & 78.72     & 56.67 \\
           &           & P+R1   & 0.01  & 49.31  & 54.52    & 64.35 & 64.48 & 83.17  & 64.81    & 79.07     & 57.54 \\
           &           & P+R5   & 0.01  & 50.73  & 56.25    & 65.20 & 67.12 & 85.13  & 67.48    & 79.86     & 59.04 \\
           &           & R1     & 0.12  & 48.91  & 47.50    & 59.53 & 74.23 & 73.19  & 46.69    & 72.21     & 54.33 \\
           &           & R5     & 0.28  & 57.96  & 57.74    & 67.44 & 77.77 & 88.13  & 68.86    & 80.14     & 65.72 \\
\cline{2-12}
           & \multirow{5}{*}{SAR}       & P      & 0.01  & 26.19  & 27.24    & 22.74 & 38.39 & 64.93  & 57.01    & 41.36     & 34.74 \\
           &           & P+R1   & 0.00  & 25.68  & 26.90    & 23.87 & 38.30 & 67.35  & 58.49    & 41.28     & 35.24 \\
           &           & P+R5   & 0.00  & 26.23  & 26.37    & 25.04 & 38.66 & 66.04  & 56.55    & 41.77     & 35.08 \\
           &           & R1     & 0.01  & 17.28  & 17.43    & 19.72 & 36.60 & 40.71  & 43.57    & 31.65     & 25.94 \\
           &           & R5     & 0.08  & 21.68  & 23.19    & 21.07 & 39.81 & 48.48  & 56.93    & 35.71     & 31.90 \\
\cline{2-12}
           & \multirow{5}{*}{SAR+Optical}   & P      & 0.00  & 49.05  & 54.34    & 63.57 & 64.59 & 83.28  & 63.28    & 78.89     & 57.18 \\
           &           & P+R1   & 0.00  & 49.15  & 54.42    & 64.81 & 65.06 & 84.33  & 62.90    & 79.53     & 57.63 \\
           &           & P+R5   & 0.01  & 50.51  & 54.86    & 65.22 & 65.90 & 84.29  & 65.86    & 80.33     & 58.47 \\
           &           & R1     & 0.07  & 50.41  & 46.20    & 59.52 & 74.11 & 75.33  & 53.95    & 72.30     & 54.82 \\
           &           & R5     & 0.30  & 57.06  & 56.39    & 66.85 & 77.00 & 89.34  & 71.42    & 80.02     & 66.05 \\
\hline \hline
\end{tabular}
}
\end{center}
\end{table*}

\begin{figure*}[!t]
    \centering
    \includegraphics[width=6.9in]{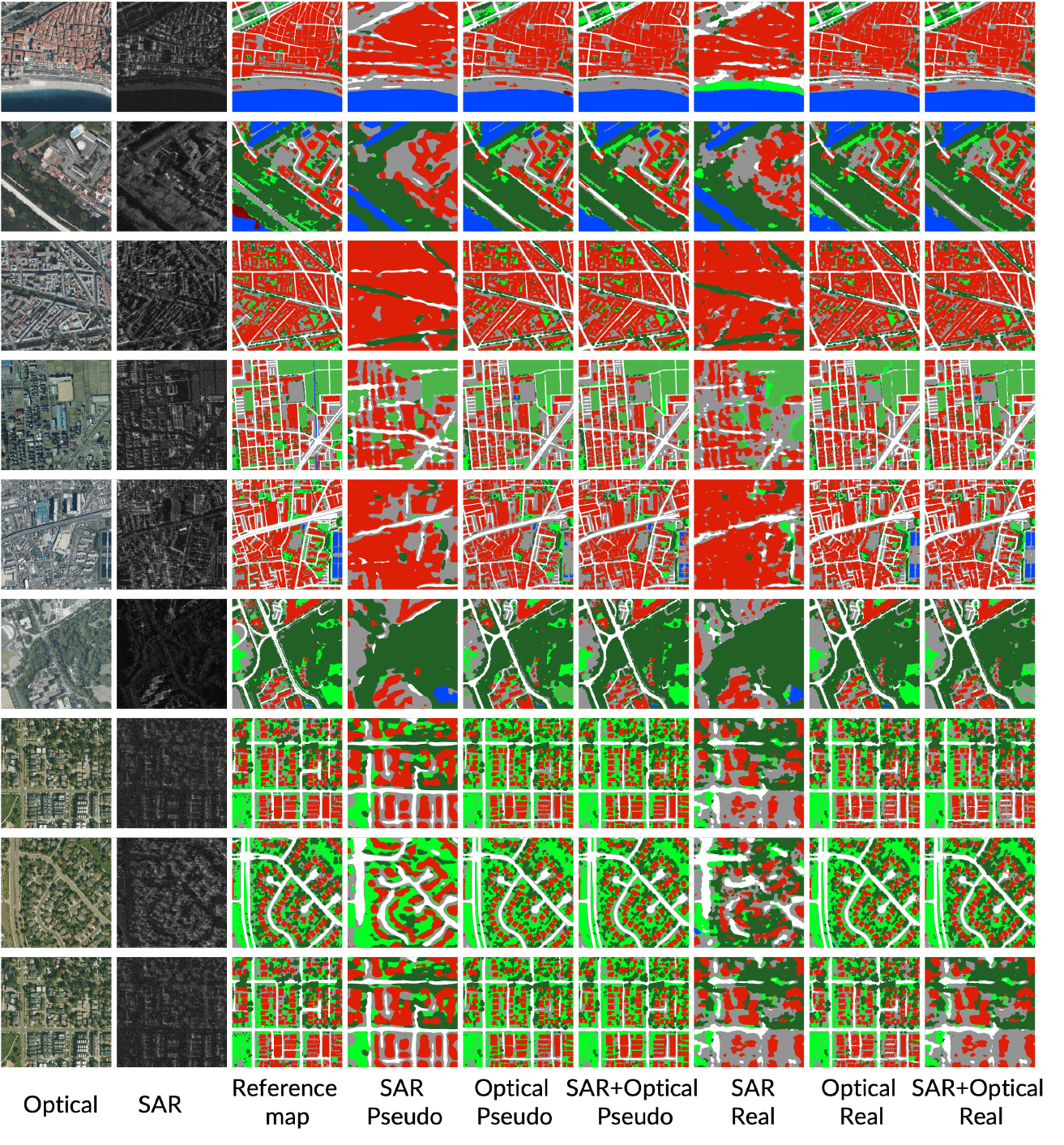}
   \caption{Visual comparison of Semantic segmentation based on U-Net with three modalities from pseudo labels (P) and real labels (R5) results as shown in Table~\ref{tab:rgb_sar_combined_results}.}
  \label{fig:classification}
\end{figure*}

\section{Land Cover Semantic Segmentation}\label{sec:3}
\subsection{Baselines}\label{sec:3.1}
For the land cover semantic segmentation task, current mainstream architectures, including those based on CNN, Transformers and Mamba, were evaluated and compared on the OpenEarthMap-SAR dataset. More specifically, the chosen models are U-Net~\cite{ronneberger2015u},  SegFormer~\cite{xie2021SegFormer}, and VMamba \cite{liu2024vmamba}. 

\subsection{Results}\label{sec:3.2}
\par In this experiment, the performance of baseline models for semantic segmentation was analyzed across different modalities: Optical, SAR, and their combination (SAR+Optical). The models evaluated include U-Net, SegFormer, and VMamba, with results presented for varying difference cases (P, P+R1, P+R5, R1, R5) in Table~\ref{tab:rgb_sar_combined_results}. Fig.~\ref{fig:classification} provide a visual comparison of semantic segmentation performance using U-Net across three modalities—Optical, SAR, and SAR+Optical—based on both pseudo labels (P) and real labels (R5). 
Across all scenarios, Optical and SAR+Optical consistently outperformed SAR alone. For VMamba, SAR+Optical provided better performance than Optical. However, for U-Net and SegFormer, SAR+Optical either matched or slightly underperformed compared to Optical.
\par For all three models, performance with Optical data improved significantly when moving from pseudo labels (P) to real labels, especially with the use of 5 real labels per region (R5). For example, U-Net's mIoU increased from 56.56\% (P) to 65.10\% (R5) with Optical data, while SegFormer and VMamba also exhibited similar improvements. This indicates that real labels are crucial for enhancing the model's segmentation accuracy, particularly for Optical-based models.
\par In contrast, while the performance of models trained on SAR data slightly improved with the inclusion of real labels in the pseudo-labels, it remained consistently lower than those trained with Optical data. Even when real labels were included, the performance did not reach the level of pseudo-labeled data. For instance, the mIoU for VMamba with SAR improved from 34.74\% (P) to 31.90\% (R5) with real labels. This indicates that SAR data, although valuable, still does not offer as much discriminative power for segmentation tasks as RGB data, particularly when only a limited number of real labels are available.
\par The combination of SAR and Optical (SAR+Optical) produced improved results compared to using SAR alone, but still did not slightly improve the performance of Optical-based models. This indicates that while combining modalities can help enhance performance, Optical remains the more effective modality for semantic segmentation tasks. The best performance was achieved using VMamba when 5 real labels per region (R5) were used (66.05\%), highlighting the importance of high-quality labeled data in driving improvements across all models.
\par The comparison of class-specific IoUs across different modalities and cases reveals several interesting trends. In general, Optical data consistently outperformed SAR, particularly for classes with distinctive color and texture features. \textit{Bareland}, \textit{Road}, and \textit{Buildings}, which exhibit strong color contrasts in Optical images, showed significantly higher IoUs when models were trained on Optical data. However, SAR data, which captures surface characteristics like roughness and reflectivity, struggled to match the performance of Optical in these classes. Despite its advantages in certain conditions, SAR data's discriminative power was limited, especially in heterogeneous environments where clear texture and color differences were crucial for segmentation.
\par When combining SAR and Optical data (SAR+Optical), the models exhibited improved performance for several classes, notably \textit{Agriculture Land} and \textit{Water}. Vegetation, which often has subtle texture differences in Optical images, benefited from the additional surface information provided by SAR. The enhanced ability to detect and segment water bodies was particularly evident in flood-prone areas, where SAR's sensitivity to moisture and the combined data's richer contextual information contributed to better IoU scores. The combination of SAR and Optical also helped improve the segmentation of urban structures, especially in regions where surface characteristics were less distinct in Optical alone.
\par Overall, the findings emphasize the substantial benefits of combining multiple modalities, such as SAR and Optical, to enhance segmentation of diverse land cover classes. While Optical data remains superior for certain features, SAR provides valuable insights, particularly for more complex environmental characteristics, especially when integrated with real or pseudo-labeled data. This synergy between the modalities suggests that future models should prioritize multi-modal approaches to achieve greater segmentation accuracy. Although SAR data proves beneficial in specific domains, it still falls short of Optical in terms of segmentation capability. The inclusion of real labels, especially in adequate quantities, significantly boosts model performance, underscoring the importance of accurate labeled datasets for training robust semantic segmentation models. In conclusion, the combination of Optical and SAR data with real labels emerges as a powerful strategy for enhancing segmentation performance and advancing geospatial analysis.

\section{Conclusion and future work}
\par We introduce OpenEarthMap-SAR, a comprehensive benchmark dataset designed for global high-resolution land cover mapping using synthetic aperture radar (SAR) data. OpenEarthMap-SAR consists of 1.5 million segments of 5033 aerial and satellite images, each with a resolution of 1024×1024 pixels, spanning 35 regions in Japan, France, and the USA. The dataset features partially manually annotated and fully pseudo-labeled 8-class land cover categories, with a ground sampling distance ranging from 0.15 to 0.5 meters. OpenEarthMap-SAR offers a unique combination of high-resolution SAR data with extensive geographic coverage, making it an invaluable resource for training and evaluating machine learning and deep learning models for land cover classification.

\par The dataset’s contribution is substantial, offering a platform for advancing land cover mapping models by providing both high-resolution imagery and diverse regional data. We present several challenges in the dataset, such as the integration of SAR data with optical imagery for more comprehensive segmentation tasks, and we propose new problem settings for further research. These challenges provide opportunities for developing more robust models capable of handling complex, real-world geospatial tasks. However, OpenEarthMap-SAR also has certain limitations. Firstly, while it includes SAR imagery, the data quality can vary due to factors like weather conditions and the inherent noise in SAR data. Furthermore, the pseudo-labels derived from optical images, while efficient, may introduce inaccuracies compared to manually labeled SAR data due to differences in data acquisition times, sensor angles, and the inherent characteristics of each modality. Variations in the timing of data collection can result in changes in the landscape not captured in both datasets. Additionally, the differences in how SAR and optical images represent objects—SAR focusing on surface structure and optical imagery capturing color and texture—can lead to mismatches. These discrepancies can contribute to label noise, making it more challenging to achieve accurate land cover classification.

\par Looking forward, there are several opportunities for expanding OpenEarthMap-SAR to further enhance its utility. The addition of other data modalities, such as height data or multispectral optical imagery, could provide complementary information to improve segmentation and classification accuracy. Incorporating more sophisticated pseudo-labeling techniques and leveraging expert manual annotations could further refine the dataset’s quality and applicability. Additionally, expanding OpenEarthMap-SAR to cover more diverse regions and types of land cover would make it even more useful for training generalized models that can be deployed in a wide range of geospatial applications.

\par OpenEarthMap-SAR has significant potential beyond land cover classification. The dataset can be used in multimodal remote sensing research, such as in land cover change detection, environmental monitoring, and disaster management. It can also support the development of remote sensing foundation models, enabling AI models that generalize across different tasks, geographies, and sensor modalities. By combining SAR with optical data, OpenEarthMap-SAR offers a foundation for building versatile models capable of improving the performance of various downstream applications in real-world geospatial settings.

\par In conclusion, OpenEarthMap-SAR represents a significant advancement in global high-resolution land cover mapping. Its ability to provide all-weather monitoring lays the foundation for diverse applications in environmental monitoring and disaster response. We anticipate that this dataset will be instrumental in developing robust, scalable models that enhance the accuracy and timeliness of geospatial analysis, offering long-term benefits for areas such as urban planning, and climate change mitigation.

\bibliographystyle{IEEEtran}
\bibliography{refs}

\begin{thebibliography}{10}
\providecommand{\url}[1]{#1}
\csname url@samestyle\endcsname
\providecommand{\newblock}{\relax}
\providecommand{\bibinfo}[2]{#2}
\providecommand{\BIBentrySTDinterwordspacing}{\spaceskip=0pt\relax}
\providecommand{\BIBentryALTinterwordstretchfactor}{4}
\providecommand{\BIBentryALTinterwordspacing}{\spaceskip=\fontdimen2\font plus
\BIBentryALTinterwordstretchfactor\fontdimen3\font minus \fontdimen4\font\relax}
\providecommand{\BIBforeignlanguage}[2]{{%
\expandafter\ifx\csname l@#1\endcsname\relax
\typeout{** WARNING: IEEEtran.bst: No hyphenation pattern has been}%
\typeout{** loaded for the language `#1'. Using the pattern for}%
\typeout{** the default language instead.}%
\else
\language=\csname l@#1\endcsname
\fi
#2}}
\providecommand{\BIBdecl}{\relax}
\BIBdecl

\bibitem{Shermeyer2020SpaceNet}
J.~Shermeyer, D.~Hogan, J.~Brown, A.~Van~Etten, N.~Weir, F.~Pacifici, R.~Hansch, A.~Bastidas, S.~Soenen, T.~Bacastow, and R.~Lewis, ``Spacenet 6: Multi-sensor all weather mapping dataset,'' in \emph{Proceedings of the IEEE/CVF Conference on Computer Vision and Pattern Recognition (CVPR) Workshops}, June 2020.

\bibitem{Adriano2021Learning}
B.~Adriano, N.~Yokoya, J.~Xia, H.~Miura, W.~Liu, M.~Matsuoka, and S.~Koshimura, ``Learning from multimodal and multitemporal earth observation data for building damage mapping,'' \emph{ISPRS Journal of Photogrammetry and Remote Sensing}, vol. 175, pp. 132--143, 2021.

\bibitem{chen2025bright}
\BIBentryALTinterwordspacing
H.~Chen, J.~Song, O.~Dietrich, C.~Broni-Bediako, W.~Xuan, J.~Wang, X.~Shao, Y.~Wei, J.~Xia, C.~Lan, K.~Schindler, and N.~Yokoya, ``Bright: A globally distributed multimodal building damage assessment dataset with very-high-resolution for all-weather disaster response,'' \emph{arXiv preprint arXiv:2501.06019}, 2025. [Online]. Available: \url{https://arxiv.org/abs/2501.06019}
\BIBentrySTDinterwordspacing

\bibitem{Michele2023High}
M.~Gazzea, A.~Solheim, and R.~Arghandeh, ``High-resolution mapping of forest structure from integrated {SAR} and optical images using an enhanced u-net method,'' \emph{Science of Remote Sensing}, vol.~8, p. 100093, 2023.

\bibitem{Chini2018Towards}
M.~Chini, R.~Pelich, R.~Hostache, P.~Matgen, and C.~Lopez-Martinez, ``Towards a 20 m global building map from sentinel-1 {SAR} data,'' \emph{Remote Sensing}, vol.~10, no.~11, 2018.

\bibitem{Hafner2022Unsupervised}
S.~Hafner, Y.~Ban, and A.~Nascetti, ``Unsupervised domain adaptation for global urban extraction using sentinel-1 {SAR} and sentinel-2 msi data,'' \emph{Remote Sensing of Environment}, vol. 280, p. 113192, 2022.

\bibitem{Michael2024Deep}
M.~Recla and M.~Schmitt, ``Deep learning-based building footprint mapping using high-resolution {SAR} data,'' in \emph{IGARSS 2024 - 2024 IEEE International Geoscience and Remote Sensing Symposium}, 2024, pp. 9983--9986.

\bibitem{TSOKAS2022117342}
A.~Tsokas, M.~Rysz, P.~M. Pardalos, and K.~Dipple, ``{SAR} data applications in earth observation: An overview,'' \emph{Expert Systems with Applications}, vol. 205, p. 117342, 2022.

\bibitem{10612244}
X.~Yang, L.~Jiao, and Q.~Pan, ``Transfer adaptation learning for target recognition in {SAR} images: A survey,'' \emph{IEEE Journal of Selected Topics in Applied Earth Observations and Remote Sensing}, vol.~17, pp. 13\,577--13\,601, 2024.

\bibitem{9351574}
X.~X. Zhu, S.~Montazeri, M.~Ali, Y.~Hua, Y.~Wang, L.~Mou, Y.~Shi, F.~Xu, and R.~Bamler, ``Deep learning meets {SAR}: Concepts, models, pitfalls, and perspectives,'' \emph{IEEE Geoscience and Remote Sensing Magazine}, vol.~9, no.~4, pp. 143--172, 2021.

\bibitem{Li2022MCANet}
X.~Li, G.~Zhang, H.~Cui, S.~Hou, S.~Wang, X.~Li, Y.~Chen, Z.~Li, and L.~Zhang, ``Mcanet: A joint semantic segmentation framework of optical and {SAR} images for land use classification,'' \emph{International Journal of Applied Earth Observation and Geoinformation}, vol. 106, p. 102638, 2022.

\bibitem{rs15112825}
X.~Lin, B.~Zhang, F.~Wu, C.~Wang, Y.~Yang, and H.~Chen, ``{SIVED}: A {SAR} image dataset for vehicle detection based on rotatable bounding box,'' \emph{Remote Sensing}, vol.~15, no.~11, 2023.

\bibitem{9761751}
P.~Zhang, H.~Xu, T.~Tian, P.~Gao, L.~Li, T.~Zhao, N.~Zhang, and J.~Tian, ``Sefepnet: Scale expansion and feature enhancement pyramid network for {SAR} aircraft detection with small sample dataset,'' \emph{IEEE Journal of Selected Topics in Applied Earth Observations and Remote Sensing}, vol.~15, pp. 3365--3375, 2022.

\bibitem{wang2023sar}
Z.~Wang, Y.~Kang, X.~Zeng, Y.~Wang, T.~Zhang, and X.~Sun, ``{SAR}-{AIR}craft-1.0: High-resolution {SAR} aircraft detection and recognition dataset,'' \emph{Journal of Radars}, vol.~12, no.~4, pp. 906--922, 2023.

\bibitem{li2024sardet100k}
Y.~Li, X.~Li, W.~Li, Q.~Hou, L.~Liu, M.-M. Cheng, and J.~Yang, ``{SARD}et-100k: Towards open-source benchmark and toolkit for large-scale {SAR} object detection,'' in \emph{The Thirty-eighth Annual Conference on Neural Information Processing Systems (NeurIPS)}, 2024.

\bibitem{9369836}
X.~Shi, S.~Fu, J.~Chen, F.~Wang, and F.~Xu, ``Object-level semantic segmentation on the high-resolution gaofen-3 fu{SAR}-map dataset,'' \emph{IEEE Journal of Selected Topics in Applied Earth Observations and Remote Sensing}, vol.~14, pp. 3107--3119, 2021.

\bibitem{9444638}
J.~Xia, N.~Yokoya, B.~Adriano, L.~Zhang, G.~Li, and Z.~Wang, ``A benchmark high-resolution gaofen-3 {SAR} dataset for building semantic segmentation,'' \emph{IEEE Journal of Selected Topics in Applied Earth Observations and Remote Sensing}, vol.~14, pp. 5950--5963, 2021.

\bibitem{ADRIANO2021132}
B.~Adriano, N.~Yokoya, J.~Xia, H.~Miura, W.~Liu, M.~Matsuoka, and S.~Koshimura, ``Learning from multimodal and multitemporal earth observation data for building damage mapping,'' \emph{ISPRS Journal of Photogrammetry and Remote Sensing}, vol. 175, pp. 132--143, 2021.

\bibitem{9150641}
J.~Shermeyer, D.~Hogan, J.~Brown, A.~Van~Etten, N.~Weir, F.~Pacifici, R.~Hänsch, A.~Bastidas, S.~Soenen, T.~Bacastow, and R.~Lewis, ``Spacenet 6: Multi-sensor all weather mapping dataset,'' in \emph{2020 IEEE/CVF Conference on Computer Vision and Pattern Recognition Workshops (CVPRW)}, 2020, pp. 768--777.

\bibitem{sun2023qqb}
Y.~Sun, Y.~Wang, and M.~Eineder, ``Quickquakebuildings: Post-earthquake {SAR}-optical dataset for quick damaged-building detection,'' \emph{IEEE Geoscience and Remote Sensing Letters}, vol.~21, pp. 1--5, 2024.

\bibitem{xia2023openearthmap}
J.~Xia, N.~Yokoya, B.~Adriano, and C.~Broni-Bediako, ``Openearthmap: A benchmark dataset for global high-resolution land cover mapping,'' in \emph{Proceedings of the IEEE/CVF Winter Conference on Applications of Computer Vision}, 2023, pp. 6254--6264.

\bibitem{yokoya2024submeter}
N.~Yokoya, J.~Xia, and C.~Broni-Bediako, ``Submeter-level land cover mapping of japan,'' \emph{International Journal of Applied Earth Observation and Geoinformation}, vol. 127, p. 103660, 2024.

\bibitem{broni2024generalized}
C.~Broni-Bediako, J.~Xia, J.~Song, H.~Chen, M.~Siam, and N.~Yokoya, ``Generalized few-shot semantic segmentation in remote sensing: Challenge and benchmark,'' \emph{IEEE Geoscience and Remote Sensing Letters}, pp. 1--5, 2024.

\bibitem{xia2023generating}
\BIBentryALTinterwordspacing
J.~Xia, C.~Broni-Bediako, and N.~Yokoya, ``Generating national very high-resolution land cover product of france without any labels: A comparative study,'' 2023. [Online]. Available: \url{https://ssrn.com/abstract=4987487}
\BIBentrySTDinterwordspacing

\bibitem{Chen2024ObjFormer}
H.~Chen, C.~Lan, J.~Song, C.~Broni-Bediako, J.~Xia, and N.~Yokoya, ``{ObjFormer: Learning Land-Cover Changes From Paired OSM Data and Optical High-Resolution Imagery via Object-Guided Transformer},'' \emph{IEEE Transactions on Geoscience and Remote Sensing}, vol.~62, pp. 1--22, 2024.

\bibitem{wang2021loveda}
J.~Wang, Z.~Zheng, A.~Ma, X.~Lu, and Y.~Zhong, ``Love{DA}: A remote sensing land-cover dataset for domain adaptive semantic segmentation,'' in \emph{Neural Information Processing Systems (NeurIPS)}, 2021.

\bibitem{demir2018deepglobe}
I.~Demir, K.~Koperski, D.~Lindenbaum, G.~Pang, J.~Huang, S.~Basu, F.~Hughes, D.~Tuia, and R.~Raskar, ``Deepglobe 2018: A challenge to parse the earth through satellite images,'' in \emph{Proceedings of the IEEE Conference on Computer Vision and Pattern Recognition Workshops}, 2018, pp. 172--181.

\bibitem{ronneberger2015u}
O.~Ronneberger, P.~Fischer, and T.~Brox, ``U-net: Convolutional networks for biomedical image segmentation,'' in \emph{International Conference on Medical image computing and computer-assisted intervention}.\hskip 1em plus 0.5em minus 0.4em\relax Springer, 2015, pp. 234--241.

\bibitem{xie2021SegFormer}
E.~Xie, W.~Wang, Z.~Yu, A.~Anandkumar, J.~M. Alvarez, and P.~Luo, ``Seg{F}ormer: Simple and efficient design for semantic segmentation with transformers,'' in \emph{NeurIPS}, 2021.

\bibitem{liu2024vmamba}
Y.~Liu, Y.~Tian, Y.~Zhao, H.~Yu, L.~Xie, Y.~Wang, Q.~Ye, and Y.~Liu, ``Vmamba: Visual state space model,'' \emph{arXiv preprint arXiv:2401.10166}, 2024.

\end{thebibliography}
\end{document}